\documentclass[fleqn,usenatbib]{mnras}

\usepackage{newtxtext,newtxmath}
\usepackage[T1]{fontenc}
\DeclareRobustCommand{\VAN}[3]{#2}
\let\VANthebibliography\thebibliography
\def\thebibliography{\DeclareRobustCommand{\VAN}[3]{##3}\VANthebibliography}

\usepackage{graphicx}	
\usepackage{amsmath}	

\newcommand\kms{km s$^{-1}$}

\newcommand\teff{$T_{\rm eff}$}
\newcommand\teffa{$T_{\rm eff,1}$}
\newcommand\teffb{$T_{\rm eff,2}$}
\newcommand\logg{$\log g$}

\newcommand\msun{M$_\odot$}
\newcommand\rsun{R$_\odot$}

\newcommand{\eb}{2M16+21}

\title[2M16+21]{Stellar properties of an actively accreting Algol-type eclipsing binary 2M16212643+2136590}

\author[M. Kounkel et al.]{
Marina Kounkel$^{1,2}$\thanks{E-mail: marina.kounkel@unf.edu},
Matteo Statti$^{3,2}$,
Avani Kulkarni$^{4}$,
Keivan G.\ Stassun$^{2}$, and Meng Sun$^{5}$
\\
$^{1}$Department of Physics and Astronomy, University of North Florida, 1 UNF Dr., Jacksonville, FL 32224, USA\\
$^{2}$Department of Physics and Astronomy, Vanderbilt University, VU Station 1807, Nashville, TN 37235, USA\\
$^{3}$Department of Physics and Astronomy, York University, 4700 Keele St, Toronto, ON M3J 1P3, Canada\\
$^{4}$Amity Regional High School, Woodbridge, CT 06525, USA\\ 
$^{5}$Center for Interdisciplinary Exploration and Research in Astrophysics (CIERA), Northwestern University, 1800 Sherman Ave,
Evanston, IL 60201, USA}

\date{Accepted XXX. Received YYY; in original form ZZZ}

\pubyear{2023}

\begin{document}
\label{firstpage}
\pagerange{\pageref{firstpage}--\pageref{lastpage}}
\maketitle

\begin{abstract}


Interacting binary stars undergo evolution that is significantly different from single stars, thus, a larger sample of such systems with precisely determined stellar parameters is needed to understand the complexities of this process. We present an analysis of a hierarchical triple containing a spectroscopically double-lined eclipsing binary, \eb. Our calculations show that this system has undergone significant mass transfer, with the current mass and radius of the donor of 0.33 \msun and 2.55 \rsun, as well as the accretor of 1.37 \msun\ and 2.20 \rsun, resulting in a mass ratio of 4.2. Despite the already significant mass loss from the donor, shedding well over half its initial gas, mass transfer remains active. The shock from the accretion has produced a spot on the surface of the accretor that is $\sim$2 times hotter than the photosphere, reaching temperatures of $\sim$10,000 K and producing significant UV excess. This shock temperature is comparable to what is seen in the pre-main sequence stars that undergo active accretion. The compactness of the hot spot of just $\sim2^\circ$ is one of the smallest observed in systems exhibiting binary mass transfer, pointing to the recency of its formation, as such it can be used to explicitly trace the point of impact of the accretion stream. The donor of this system may be a sub-sub-giant; comparing it with systems with similar initial conditions may help with understanding the formation processes of such stars.

\end{abstract}



\section{Introduction}

Eclipsing binaries (EBs) are valuable laboratories for understanding stellar evolution, as they offer a direct method probe of masses and radii of the individual stars that is independent of any model-dependent assumption. However, evolution of such systems can also be fundamentally different than in single stars. In particular, close binary systems can undergo transfer of mass from one star onto another if one of the stars fills out its Roche lobe, the size of which depends on the proximity of the two stars and their mass ratio. The resulting distortion of these stars can be observed in the light curves of EBs. If only one star fills out its Roche lobe, such systems are typically referred to as semi-detatched or Algol-type eclipsing binaries.

Mass transfer in binary systems can occur at different states of stellar evolution. Case A includes systems consisting of main sequence stars, Case B involves systems that have begun following exhaustion of H in the core and transition of the star into the red giant phase, and Case C following exhaustion of He \citep{ivanova2015}. Similarly, it is possible to catch systems at different times from the onset of mass transfer. Depending on this, mass accretion rates could vary wildly, ranging from $10^{-4}$ to $10^{-12}$ \msun\ year${^{-1}}$ \citep{van-rensbergen2011,manzoori2017}.

In order to understand these complex processes, a sample of multiple systems consisting of stars of different masses and evolutionary stages with precisely determined stellar parameters is needed. 

Previously, we have modelled 2M17091769+3127589, a double-lined eclipsing binary system \citep{miller2021a}. This system was notable for having the more luminous star to have mass almost 6 times smaller than its fainter companion, which can be a byproduct of mass transfer from an evolved giant onto a main sequence star. The initial characterization of this system was serendipitous: it was one of the few eclipsing binaries identified in the ASAS-SN survey at the time \citep{pawlak2019} for which there were a sufficient number of high resolution spectra taken at different dates in which the star appeared as a double lined spectroscopic binary (SB2) that enabled to measure radial velocities (RVs) of both stars.

In this paper, we perform modeling of a system that is a twin to 2M1709, originating with stars of similar masses and comparable separation, but caught in the earlier stage of its evolution. This system, 2M16212643+2136590 (hereafter \eb), is a semi-detatched eclipsing binary, and it still appears to have a significant ongoing accretion. In Section \ref{sec:data} we describe the available photometric, spectroscopic, and astrometric data that are currently available towards \eb. In Section \ref{sec:model} we model the physical and orbital properties of \eb, including the masses and radii of the individual stars. In Section \ref{sec:discussion} we analyze the accretion signatures, and construct a model of its evolutionary history. Finally, in Section \ref{sec:conclusions} we conclude our findings.

\section{Data} \label{sec:data}

\subsection{Radial velocity}

APOGEE is a multi-object spectrograph with high resolution (R$\sim$22,500) covering the wavelength range of 1.51---1.7 $\mu m$ \citep{majewski2017}. \citet{kounkel2021} have recently performed a systematic search of SB2s in the APOGEE data. They were identified using Gaussian fitting to the resulting cross-correlation function to measure both the RVs of the individual stars. When multiple spectral visits were available, mass ratio ($q$) was also estimated using the resulting RVs. This makes it possible to identify unusual systems, such as those with $q>2$, which signifies brighter star in the system being significantly less massive. Although there are a number of such candidates, only 3 appear to be eclipsing binaries in TESS data, and only one, \eb, has a sufficient number of APOGEE epochs to fully sample the RV phase space.

We adopt RVs for \eb\ from \citet{kounkel2021}. A total of 5 epochs are available, observed between June 2016 and February 2017. The source can be resolved as an SB2 during 3 of these epochs. Both sources are characterized by a wide profile attributable to fast rotation, with the full width of half maximum (FWHM) of the cross-correlation function of 58 \kms for the more massive star, and 68 \kms\ for the less massive star.

\begin{figure}
\includegraphics[width={0.48\textwidth}]{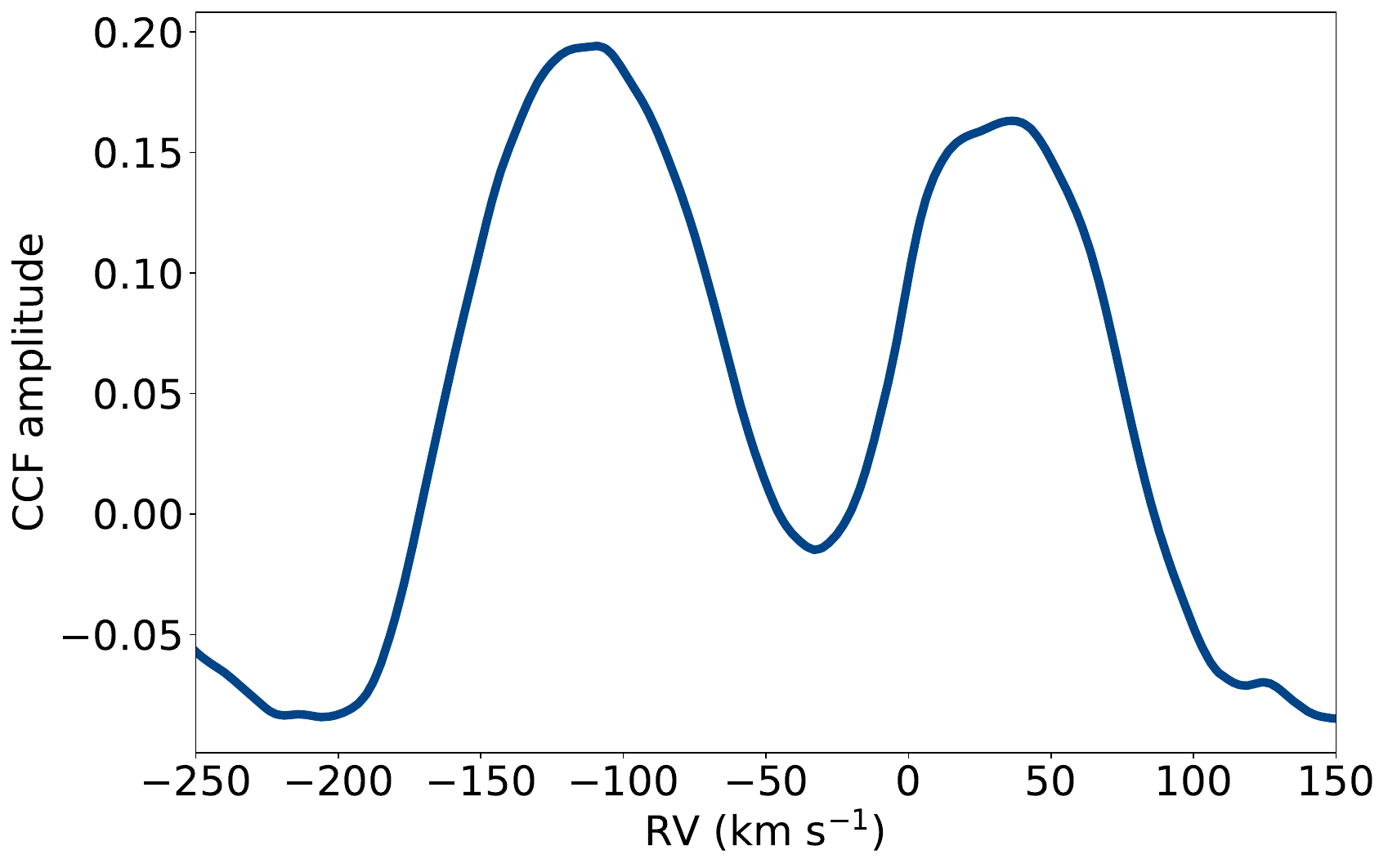}
\caption{APOGEE cross-correlation function of \eb\ showing both stars. The more pronounced blueshifted peak corresponds to the lower mass star. FWHM of the two components are 68 and 58 \kms.
\label{fig:ccf}}
\end{figure}

\begin{figure*}
\includegraphics[width={0.48\textwidth}]{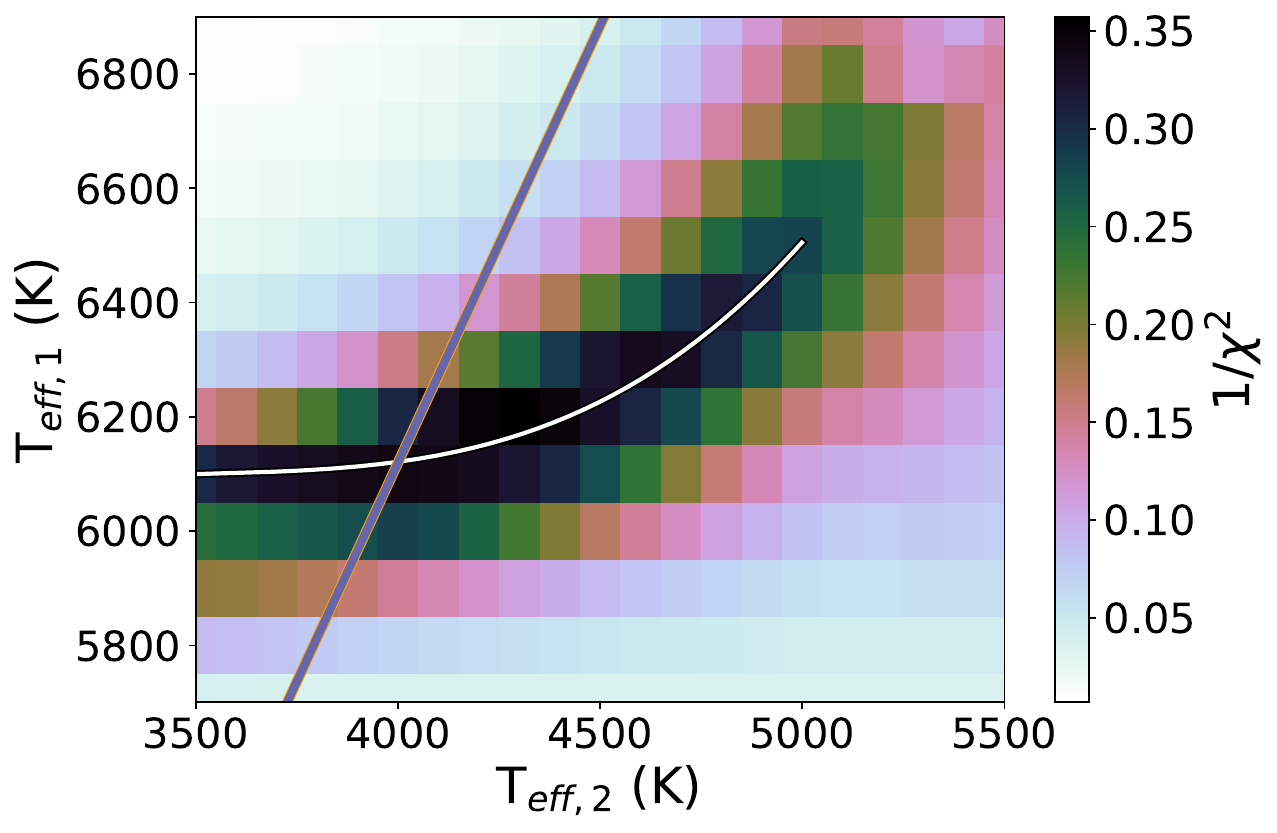}
\includegraphics[width={0.48\textwidth}]{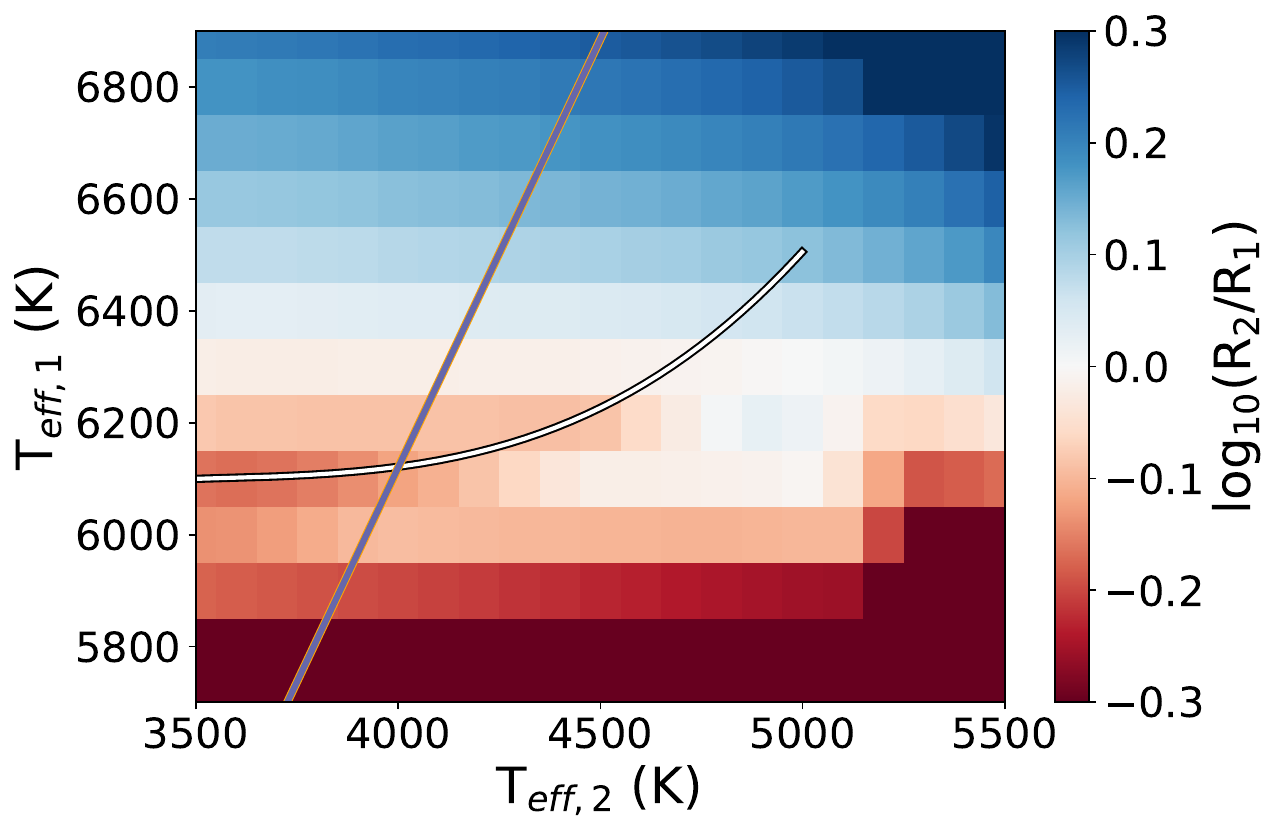}
\caption{Left: Goodness of fit of the SED using a combination of two model atmospheres of different \teff. The white line shows the best family of fits. The purple line shows the \teff\ ratio of 1.53 derived from the light curve analysis. Right: radius ratio of the two stars produced by the best fit for a given \teff\ combination.
\label{fig:sedparams}}
\end{figure*}

\begin{table*}
	\centering
	\caption{Table
\label{tab:params}}
	\begin{tabular}{cccc} 
		\hline
		Parameter & Spotless & Spotted & Fixed \teff\\
		\hline
 $P$ (day) & 2.662564683$^{+6.6e-08}_{-6.3e-08}$ & 2.662564657$^{+6.1e-08}_{-6.5e-08}$ & 2.66256467$^{+6.2e-08}_{-6.4e-08}$ \\
 $a$ (\rsun) & 9.501$^{+0.097}_{-0.101}$& 9.573$^{+0.091}_{-0.096}$ & 9.646$^{+0.093}_{-0.096}$ \\
 $i$ ($^\circ$) & 83.416$^{+0.015}_{-0.015}$ & 83.471$^{+0.014}_{-0.017}$ & 83.489$^{+0.016}_{-0.017}$ \\
 $q$ & 0.24937$^{+0.00037}_{-0.00037}$ & 0.24821$^{+0.00039}_{-0.00037}$ & 0.24169$^{+0.00039}_{-0.00038}$ \\
 $t_0$ (MJD) & 58986.473002$^{+2.1e-05}_{-2e-05}$ & 58986.472661$^{+2.4e-05}_{-2.1e-05}$ & 58986.473074$^{+1.9e-05}_{-2e-05}$ \\
 $\gamma$ (\kms) & 4.7$^{+0.6}_{-0.62}$ & 4.8$^{+0.59}_{-0.46}$ & 4.57$^{+0.6}_{-0.6}$ \\
 		\hline
 $R_1$ (\rsun) & 2.174$^{+0.022}_{-0.023}$ & 2.129$^{+0.021}_{-0.021}$ & 2.202$^{+0.021}_{-0.022}$ \\
 $R_2$ (\rsun) & 2.533$^{+0.025}_{-0.03}$ & 2.549$^{+0.025}_{-0.028}$ & 2.549$^{+0.025}_{-0.024}$\\
 $M_1$ (\msun) & 1.299$^{+0.04}_{-0.041}$ & 1.33$^{+0.038}_{-0.039}$ & 1.368$^{+0.04}_{-0.04}$ \\
 $M_2$ (\msun) & 0.324$^{+0.01}_{-0.01}$ & 0.3303$^{+0.0094}_{-0.0099}$ & 0.3306$^{+0.0098}_{-0.0097}$ \\
 \teffa\ (K) & 5872.0$^{+15.0}_{-84.0}$ & 5827.0$^{+47.0}_{-47.0}$ & 6150\\
 \teffb\ (K) & 3744.2$^{+9.1}_{-54.4}$ & 3799.0$^{+26.0}_{-26.0}$ & 3885.1$^{+1.2}_{-1.2}$\\
 \logg$_1$ (dex) & 3.8773$^{+0.0045}_{-0.0047}$ & 3.9057$^{+0.0041}_{-0.0044}$ & 3.8886$^{+0.0043}_{-0.0044}$\\
 \logg$_2$ (dex) & 3.1417$^{+0.0042}_{-0.0044}$ & 3.1444$^{+0.0043}_{-0.0041}$ & 3.1446$^{+0.0043}_{-0.0039}$ \\
 \hline
 $l_{3,{\rm T}}$ & 0.12211$^{+0.00033}_{-0.00035}$ & 0.12201$^{+0.00054}_{-0.00054}$ & 0.1229$^{+0.00034}_{-0.00034}$ \\
 $l_{3,{\rm G}}$ & 5.8e-05$^{+9.8e-05}_{-4.4e-05}$ & 0.00029$^{+0.00047}_{-0.00022}$ & 7.5e-05$^{+0.000121}_{-5.6e-05}$ \\
 $l_{3,{\rm RP}}$ & 0.02499$^{+0.00064}_{-0.00067}$ & 0.02405$^{+0.00083}_{-0.00086}$ & 0.02625$^{+0.00064}_{-0.00065}$ \\
 $l_{3,{\rm BP}}$ & 6.2e-05$^{+9.7e-05}_{-4.5e-05}$ & 0.00306$^{+0.00091}_{-0.00083}$ & 4.7e-05$^{+7.6e-05}_{-3.6e-05}$ \\
 $l_{3,{\rm V}}$ & 0.0654$^{+0.0012}_{-0.0012}$ & 0.0709$^{+0.0014}_{-0.0013}$ & 0.0638$^{+0.0011}_{-0.0012}$ \\
 $l_{3,{\rm g}}$ & 0.05879$^{+0.00073}_{-0.0007}$ & 0.06735$^{+0.00096}_{-0.00103}$ & 0.05668$^{+0.00069}_{-0.0007}$ \\
\hline
 Temp. contrast & & 2.585$^{+0.032}_{-0.036}$ & 1.8\\
 Spot size ($^\circ$) & & 3.667$^{+0.113}_{-0.094}$ & 2.2\\
 $l$ ($^\circ$) & & 131.89$^{+0.37}_{-0.37}$ & 108.52$^{+0.31}_{-0.3}$ \\
 $b$ ($^\circ$) & & 83.9$^{+0.18}_{-0.16}$ & 6.52$^{+0.66}_{-0.67}$ \\

		\hline
	\end{tabular}
\end{table*}

\begin{figure*}
\includegraphics[width={1\textwidth}]{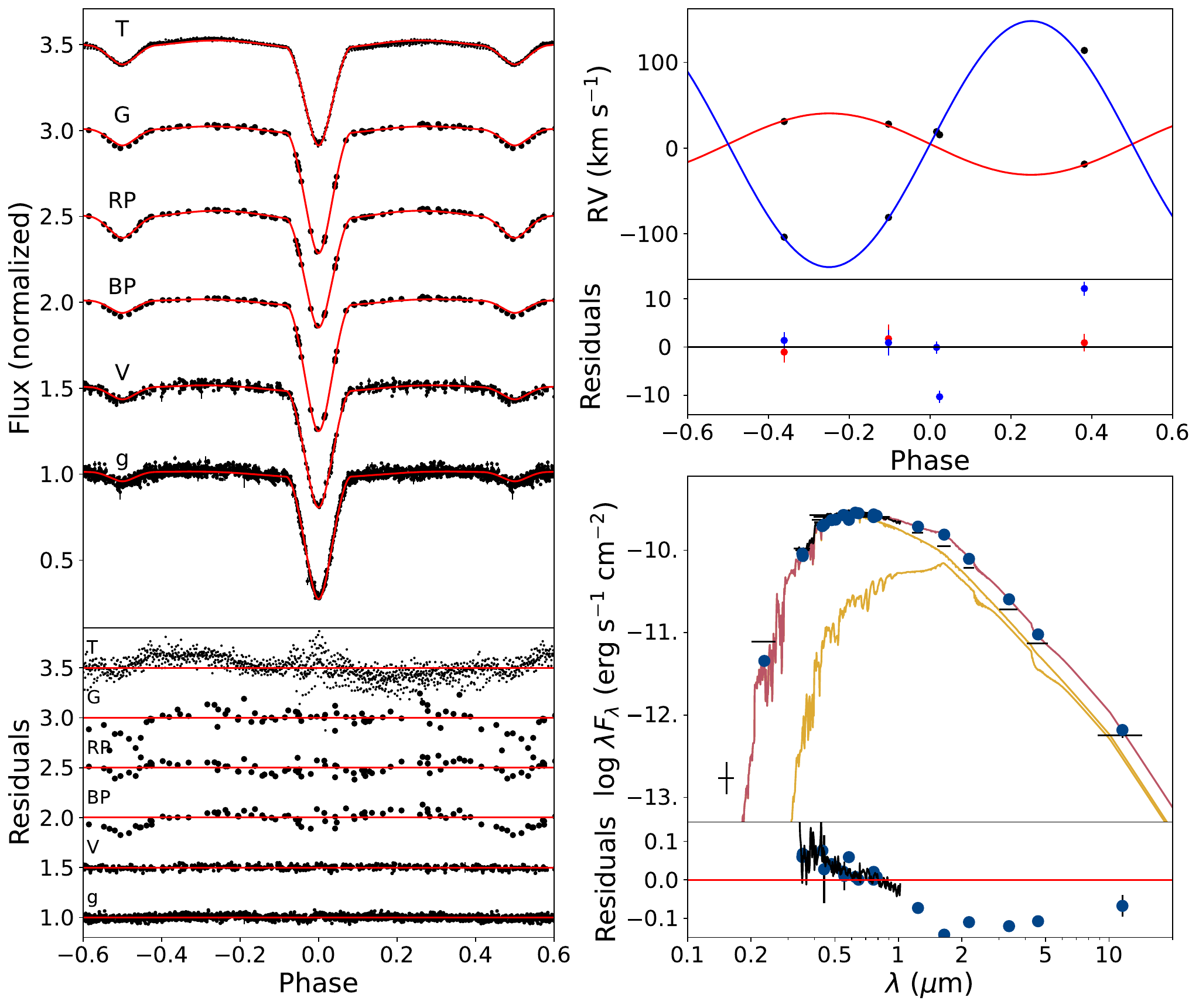}
\caption{Fit of the spotless model. Left panel shows the 6 different light curves used in the analysis, with the red line being the model, and black points being the phase-folded data. Top right panel shows the radial velocity curves for the primary and the secondary. Bottom right panel shows the SED fit, with black points corresponding to the broad-band photometry, black curve showing the Gaia XP spectrum, yellow curves showing the best-fitted SEDs of individual stars, red is the coadded SED, and blue dots show the interpolated fluxes of the model at the bandpasses corresponding to the data. Residuals between the model and the data are shown below; for light curves residuals are scaled proportionally to their respective uncertainties.
\label{fig:withoutspot}}
\end{figure*}

\subsection{Light curves}

\eb\ has a light curve morphology of an Algol-type EB with the period of $\sim$2.6 days. It has been observed by TESS in Sectors 25 (May 2020), and 51 (April 2022). However, Sector 51 has a significant portion of the data flagged as anomalous. As during a single sector, \eb\ is able to complete a full orbit 10 times, the light curve is well-sampled even with just one month of observations, as such we restricted our analysis solely to Sector 25.

The light curve was generated from TESS Full Frame Images using {\tt eleanor} \citep{feinstein2019}. The aperture was manually defined across 4 pixels that have shown the strongest periodic signal consistent with the source. We used {\tt CORR\_FLUX}, as it produced the cleanest light curve without any systematic trends and only minimal aberrations that were manually removed.

However, \eb\ is located in a somewhat crowded field. A nearby source, TIC 458723453 is located 8'' away, which is not resolved in TESS due to its large pixel scale of 21''. Modeling the pixel response function of all of the field stars in the region using Gaia photometry, we estimate that as much as 10\% of all flux in the aperture is contamination, which is important to consider in the modeling of the lightcurve.

We extract ASAS-SN light curve for \eb. It has been observed from February 2013 through August 2018 in V band, and from September 2017 onward in g band. Both light curves show similar morphology to the one in TESS, without any long term trends. However, some isolated epochs (most prominently in V band) show excessive level of noise, they were manually excluded from the analysis. Contamination from TIC 458723453 may also be a concern for ASAS-SN, as its pixel scale is 8'' \citep{kochanek2017}.

Finally, epoch photometry for \eb\ is also available as a part of Gaia DR3 \citep{eyer2022}, in $G$, $BP$, and $RP$ bands. Although only 71 epochs are available, they provide a comprehensive phase coverage of the light curve folded by the period. We also include these data in the analysis, as they are able to provide unblended fluxes unaffected by the nearby field stars.

\subsection{Stellar properties}

Parallax of 0.7485$\pm$0.1031 mas for \eb\ was reported in Gaia DR3, however it also has an extremely high renormalized unit weight error (RUWE) of 6.482. Indeed, in Gaia DR2, parallax was reported to be 0.0906$\pm$0.1198 mas, showing significant evolution with more data, as such either of these distance measurements are unlikely to be reliable. RUWE$>1.4$ are considered as spurious astrometric solutions, typically thought to be affected by the orbital motion of multiple systems \citep{gaia-collaboration2021}. However, the eclipsing binary itself is unlikely to be responsible for high RUWE: given the compactness of the orbit their orbital motion is unlikely to cause significant astrometric jitter. Rather, we suspect that this system is likely to be a hierarchical triple, housing a fainter and more widely separated companion.

As the system is a spectroscopic binary, spectroscoically derived properties are likely to be imprecise, but, nonetheless, they offer a sense of likely parameter ranges. Different pipelines analyzing APOGEE data report \teff\ ranging between 4500 and 5400 K \citep[e.g.][]{ting2019, olney2020, sprague2022, abdurrouf2022}, with the large scatter driven by different pipelines being sensitive to various features associated with both the primary and the secondary. Optical spectra for \eb\ have also been obtained by LAMOST, providing somewhat more stable \teff\ of 6240--6280 K \citep{luo2019,xiang2019,luo2019}, most likely primarily sensitive towards the hotter star. Typically reported \logg\ range between 3.5--4 dex. There is a significant scatter in the reported metallicitities, -0.1--0 dex from APOGEE, and 0.3--0.5 dex from LAMOST spectra.


We can achieve stricter constraints on \teff\ of the stars through performing an spectral energy distribution (SED) fit to the system considering two stars\footnote{\url{https://github.com/mkounkel/SEDFit}, \citet{sedfit}}. We follow the procedure outlined in \citet{stassun2016a}. Using synthetic stellar atmospheres from \citet{kurucz1993}, we pick two templates of a given temperature, with \logg=3.5 and solar metallicity. The combination of these atmospheres are evaluated against the SED of the system, consisting of low resolution XP spectrum from Gaia, which spans the range in wavelength of 336---1020 nm, as well as the fluxes in Johnson and Cousins filters, alongside with fluxes 2MASS, WISE, SDSS, Gaia, and GALEX surveys. However, we include only NUV flux from GALEX, as FUV shows excess relative to the model in all of the fits. 

We solve for radius of individual stars, distance and the extinction along the line of sight, limited to the maximum of $A_V$=0.153 mag, which is the integrated $A_V$ along this line of sight from the map of \citet{schlegel1998}. Each combination of temperatures of the atmospheres is evaluated independently, and the lowest $\chi^2$ of the fit is recorded. The best fit was produced by \teffa=6200 K, \teffb=4300 K, and angular radii of 14.0 and 11.4 $\mu$as respectively. We note that this \teffa\ well matches the spectroscopically derived \teff\ from the optical spectra, and \teffb\ is somewhat consisted with the near-IR spectra. A family of similar solutions is also possible (Figure \ref{fig:sedparams}). The fit suggests that the hotter star has to have a larger radius than the cooler star for all \teffb<6300 K.

\section{Model} \label{sec:model}

\begin{figure}
\includegraphics[width={0.48\textwidth}]{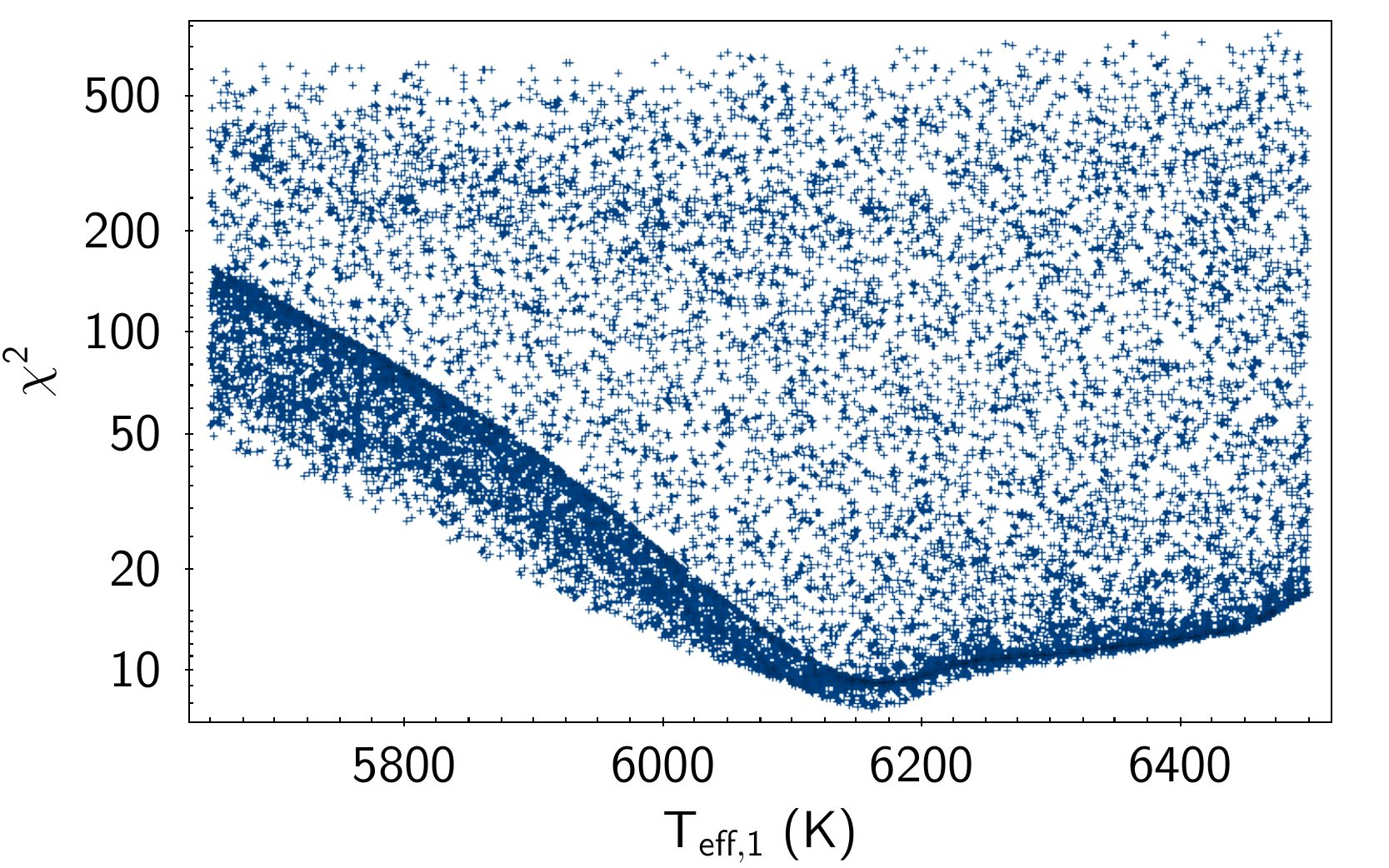}
\caption{Goodness of fit of the SED of the spotted model based on \teffa.
\label{fig:sedspot2}}
\end{figure}

\begin{figure}
\includegraphics[width={0.48\textwidth}]{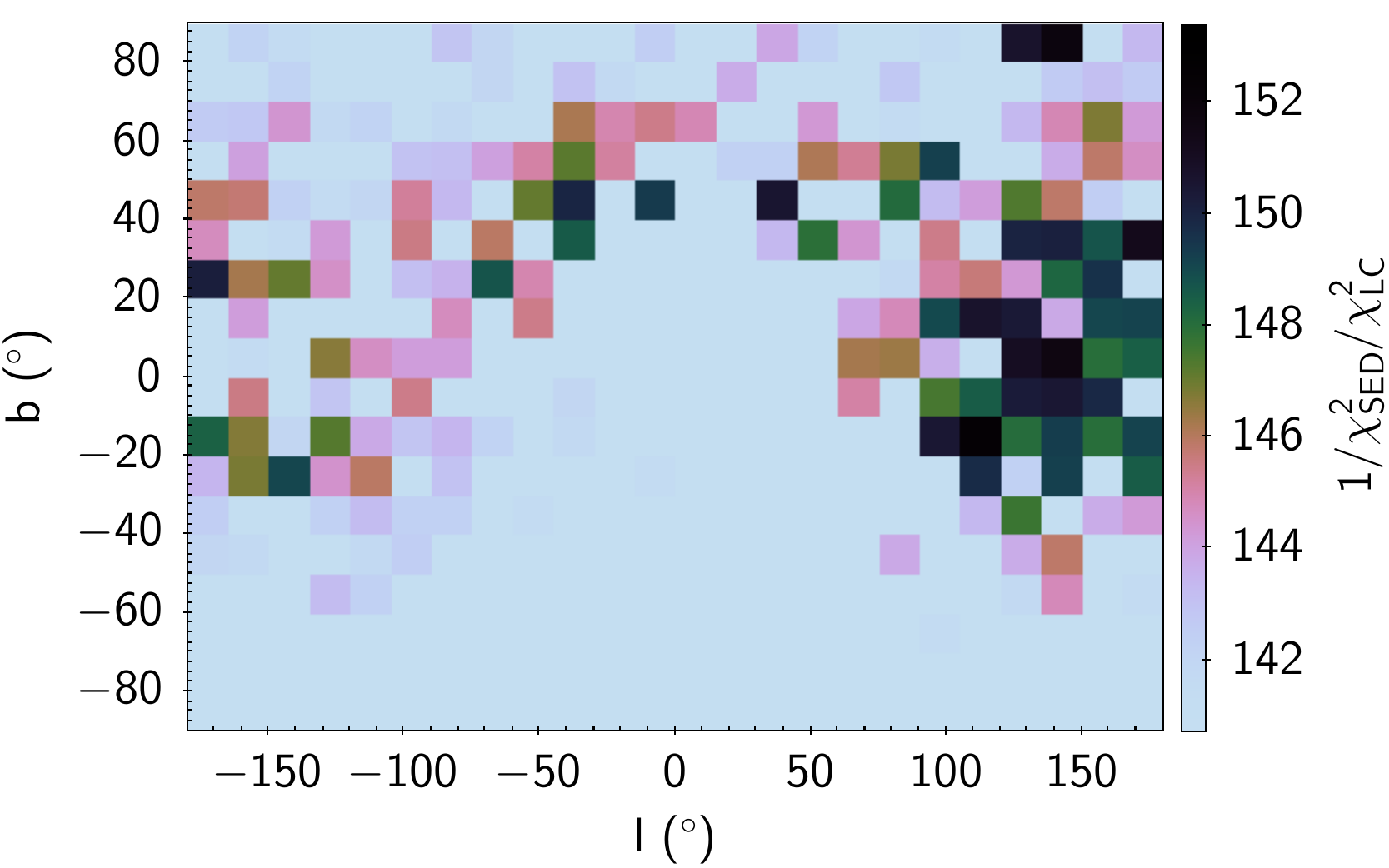}
\caption{Likelihood of the position of the star spot on the based on the goodness of fit of the spotted model using the joint solution from the SED and the light curves.
\label{fig:sedspot}}
\end{figure}

\begin{figure}
\includegraphics[width={0.48\textwidth}]{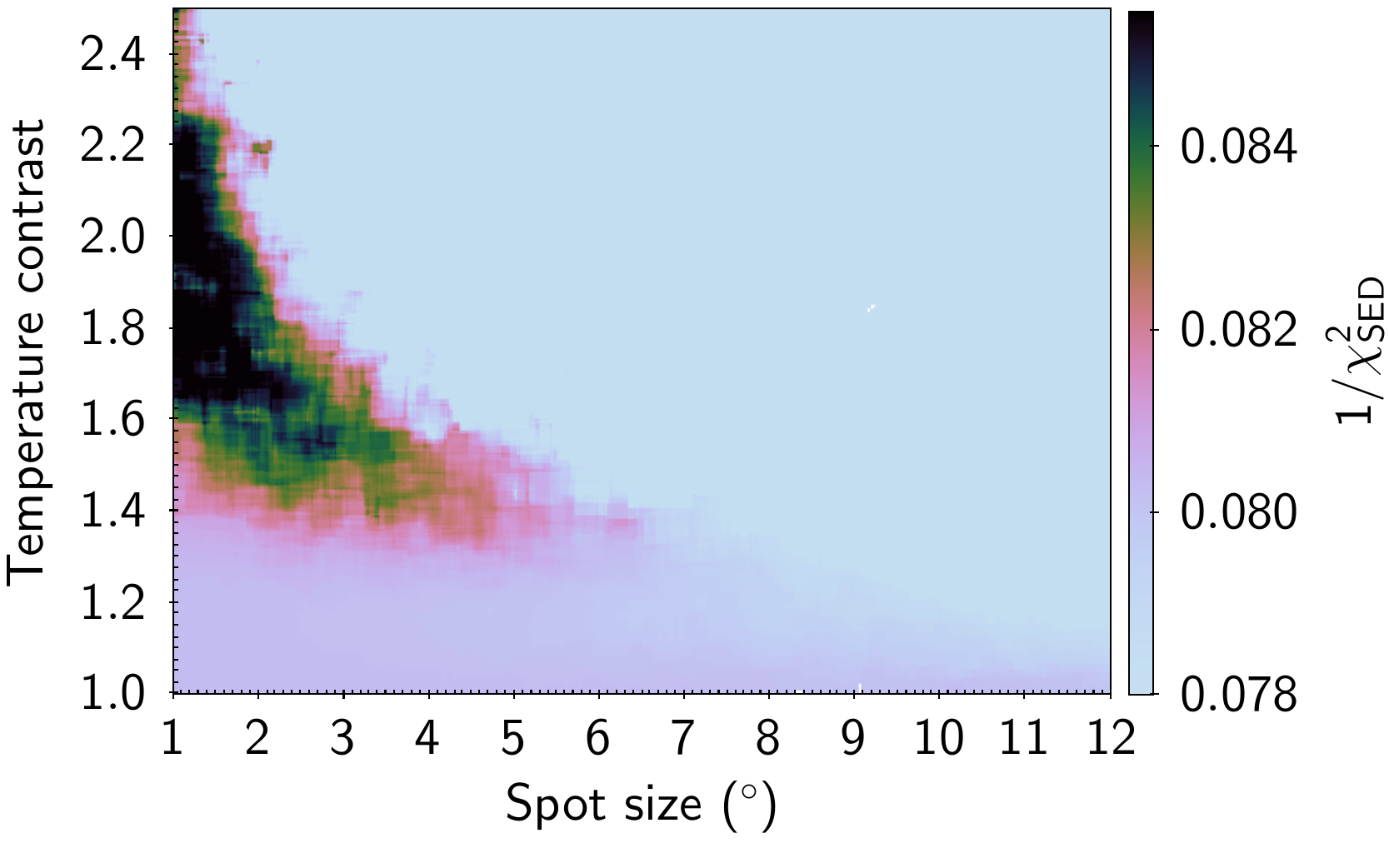}
\includegraphics[width={0.48\textwidth}]{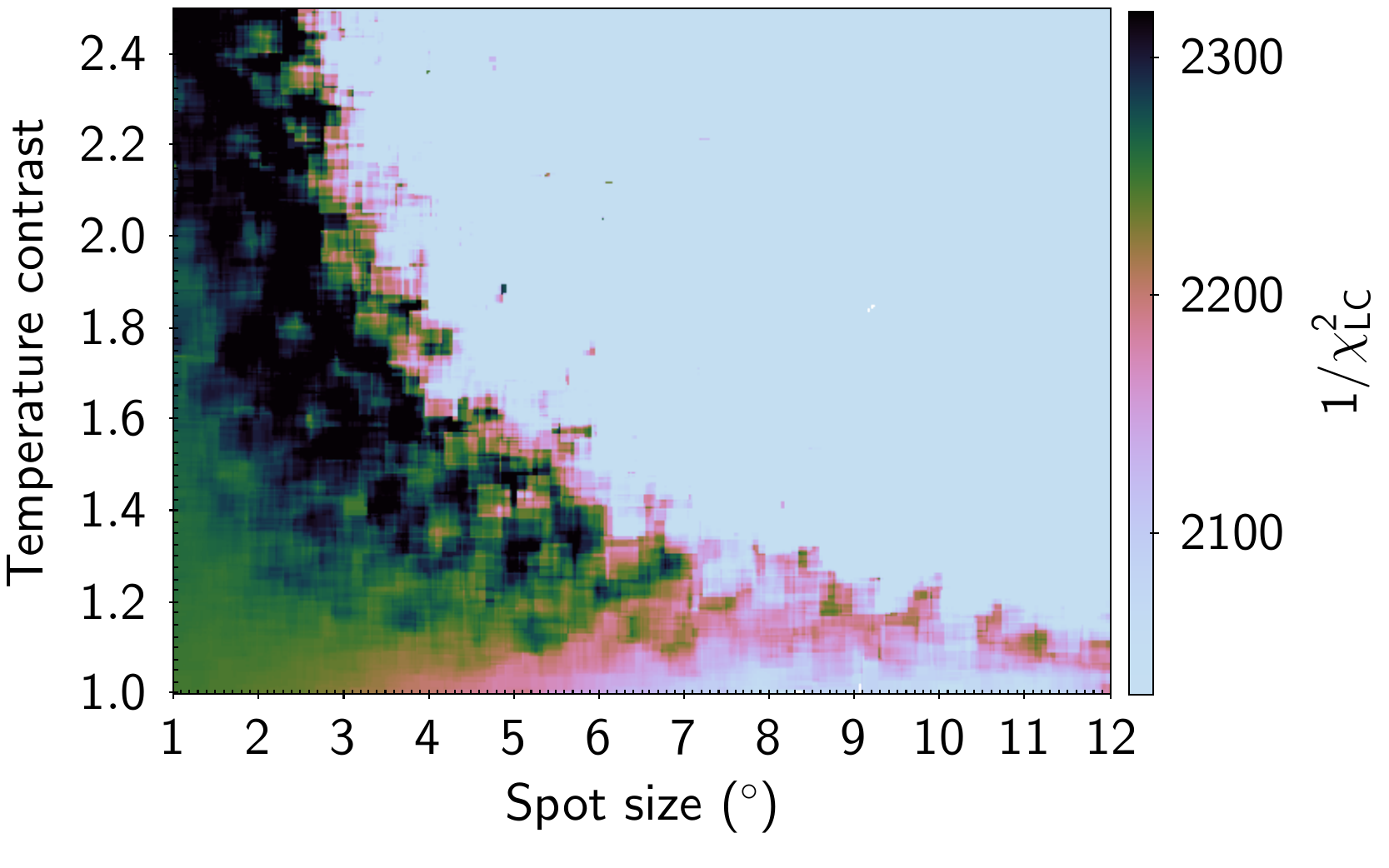}
\includegraphics[width={0.48\textwidth}]{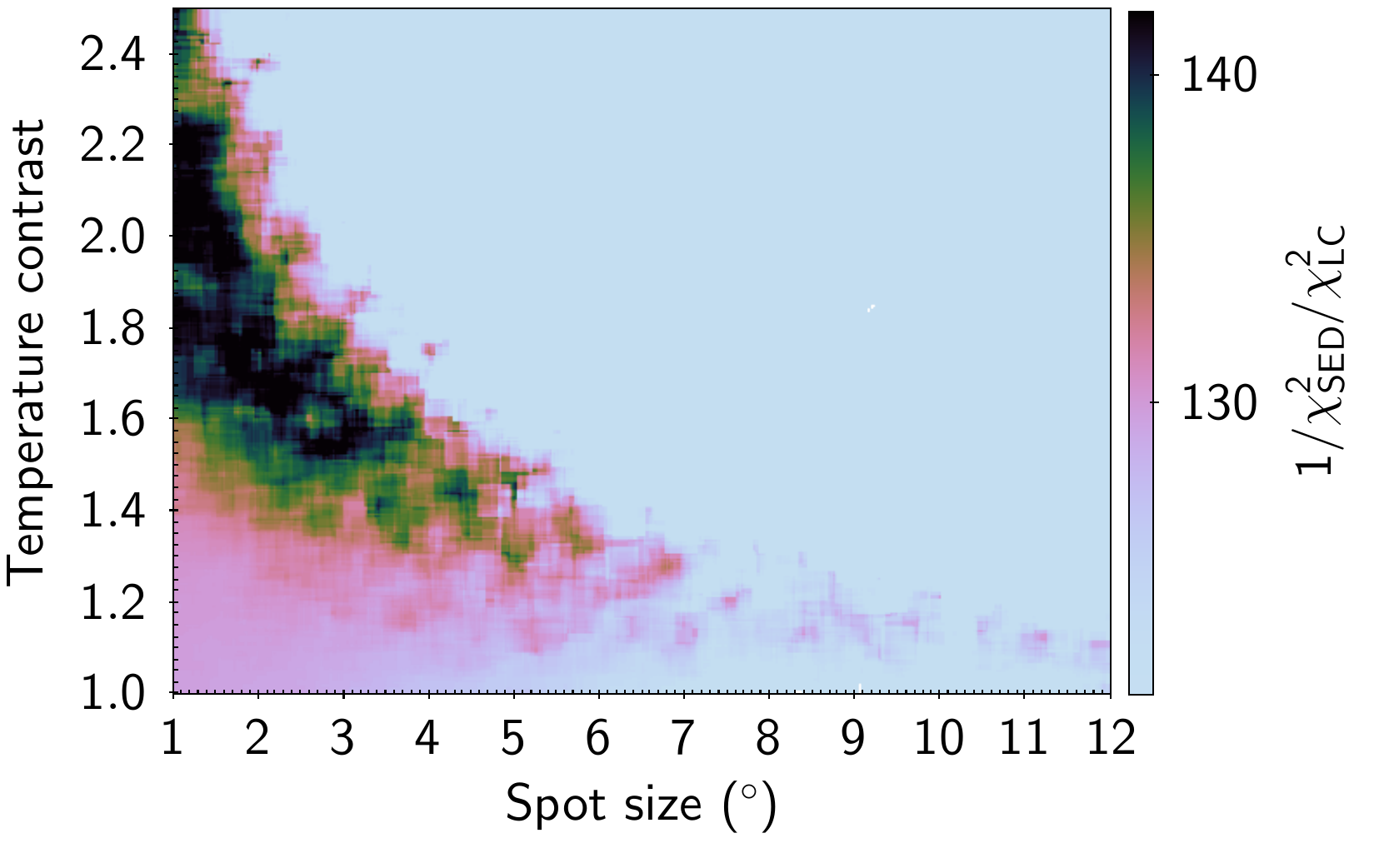}
\caption{Goodness of fit of the spottded model with variable hot spot size and temperature contrast. Top: SED-only fit. Middle: Light curve only fit. Bottom: Joint solution.
\label{fig:sedspot1}}
\end{figure}

\begin{figure*}
\includegraphics[width={1\textwidth}]{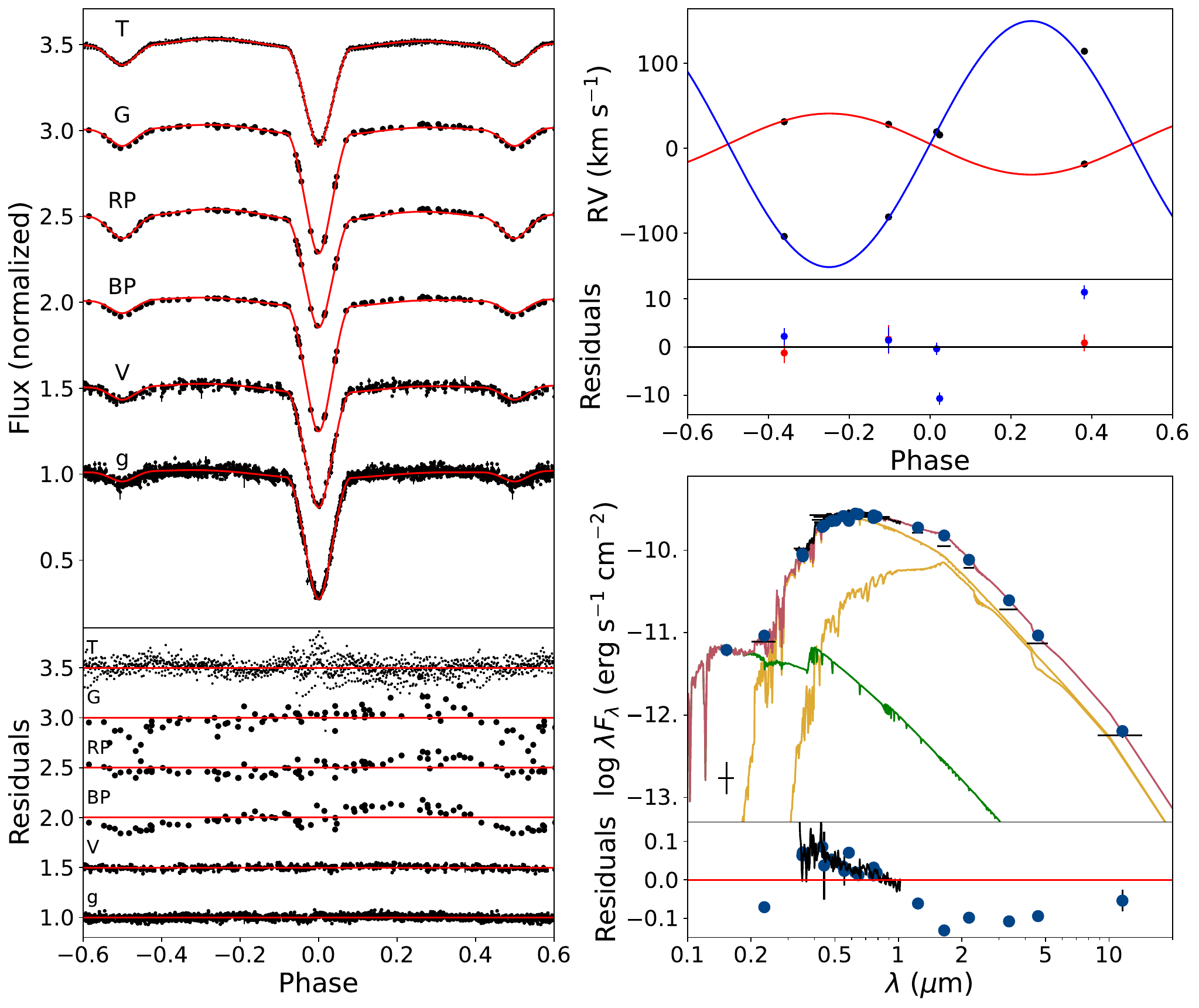}
\caption{Same as Figure \ref{fig:withoutspot}, but for a model that considers a presence of a hot spot; green curve in the SED panel is the spot contribution. Note that the fit for T band is improved over Figure \ref{fig:withoutspot}, at a cost of significant overestimation of UV flux.
\label{fig:withspot}}
\end{figure*}

\begin{figure*}
\includegraphics[width={1\textwidth}]{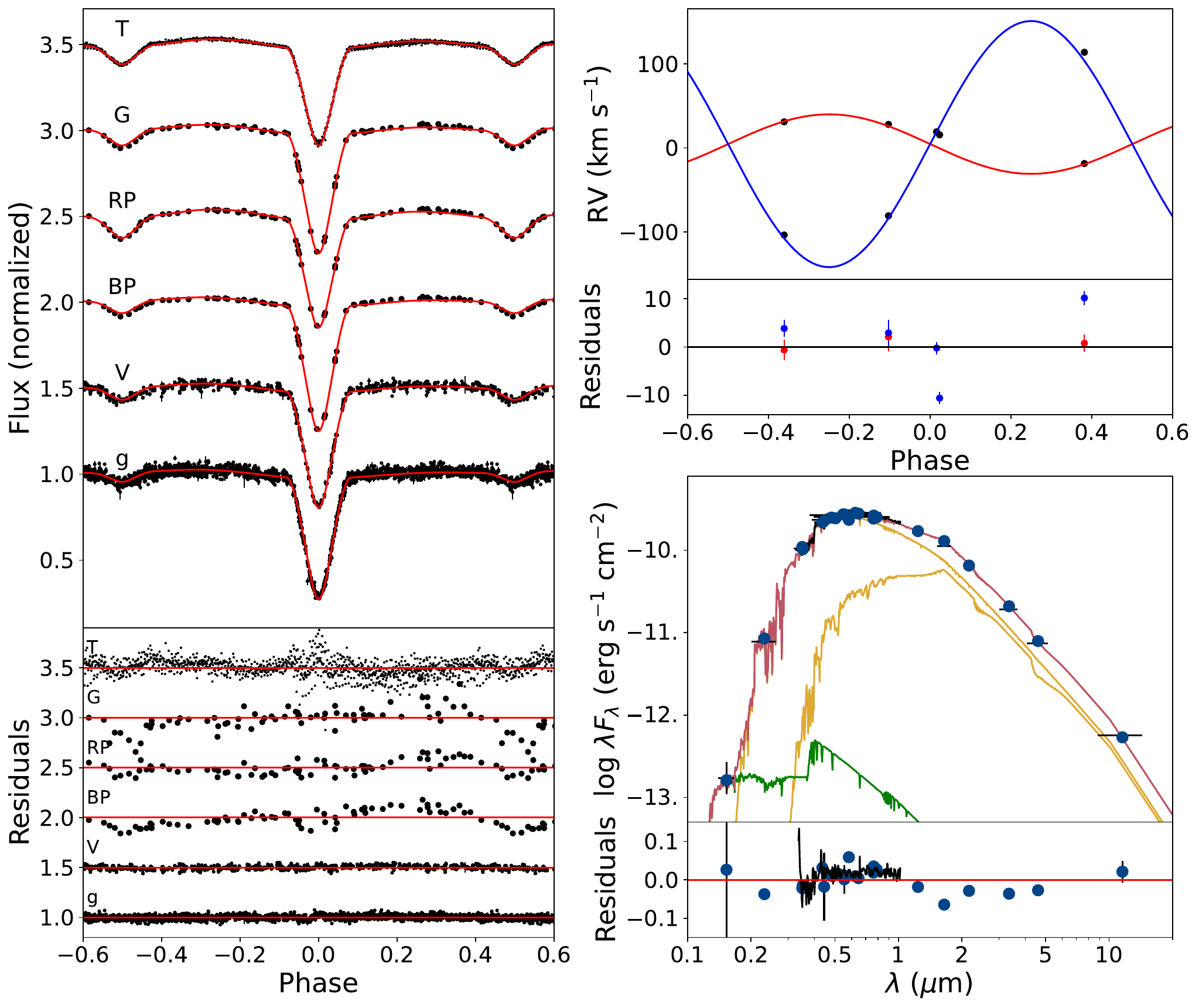}
\caption{Same as Figure \ref{fig:withspot}, but with fixed temperatures of both the spot and the photospheres. 
\label{fig:withspotfix}}
\end{figure*}

\begin{figure*}
\includegraphics[width={1\textwidth}]{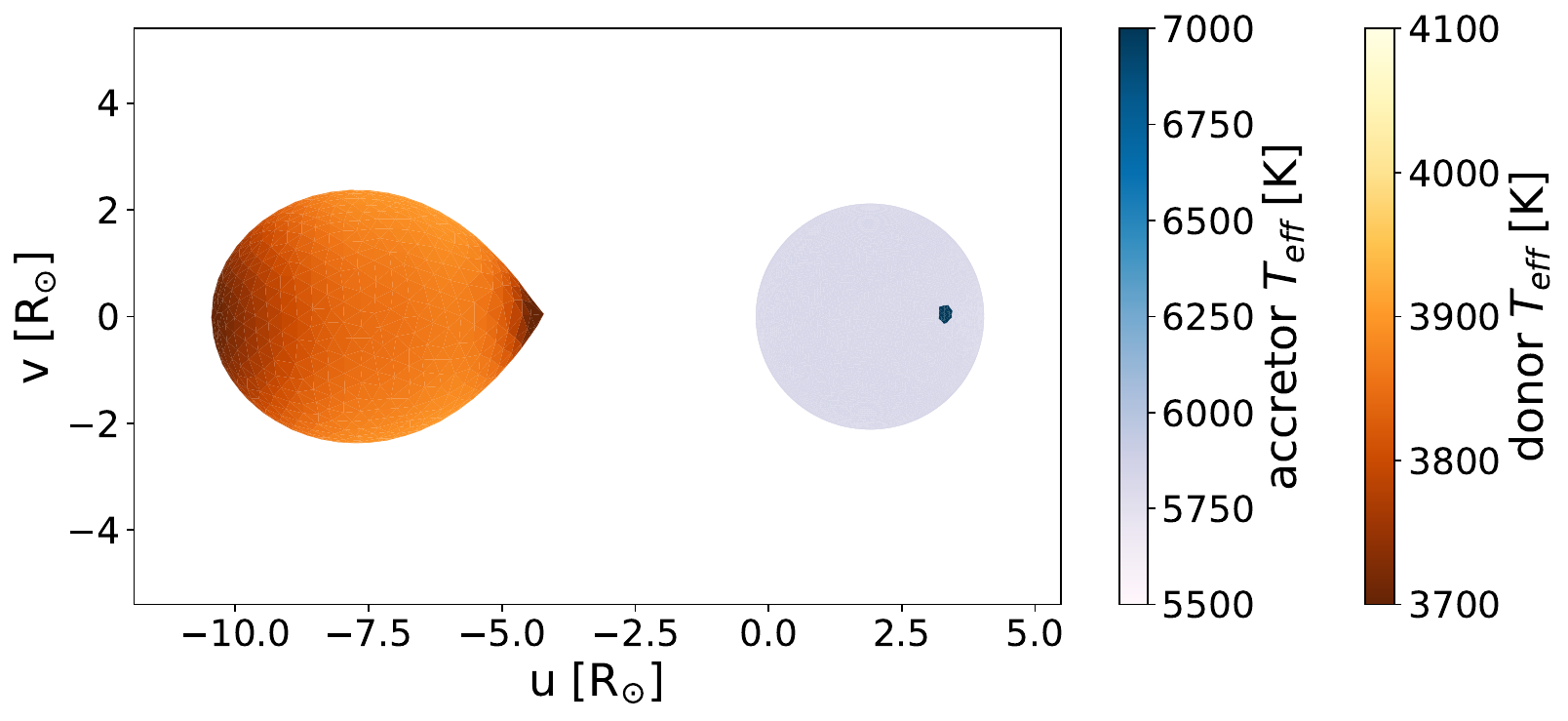}
\includegraphics[width={1\textwidth}]{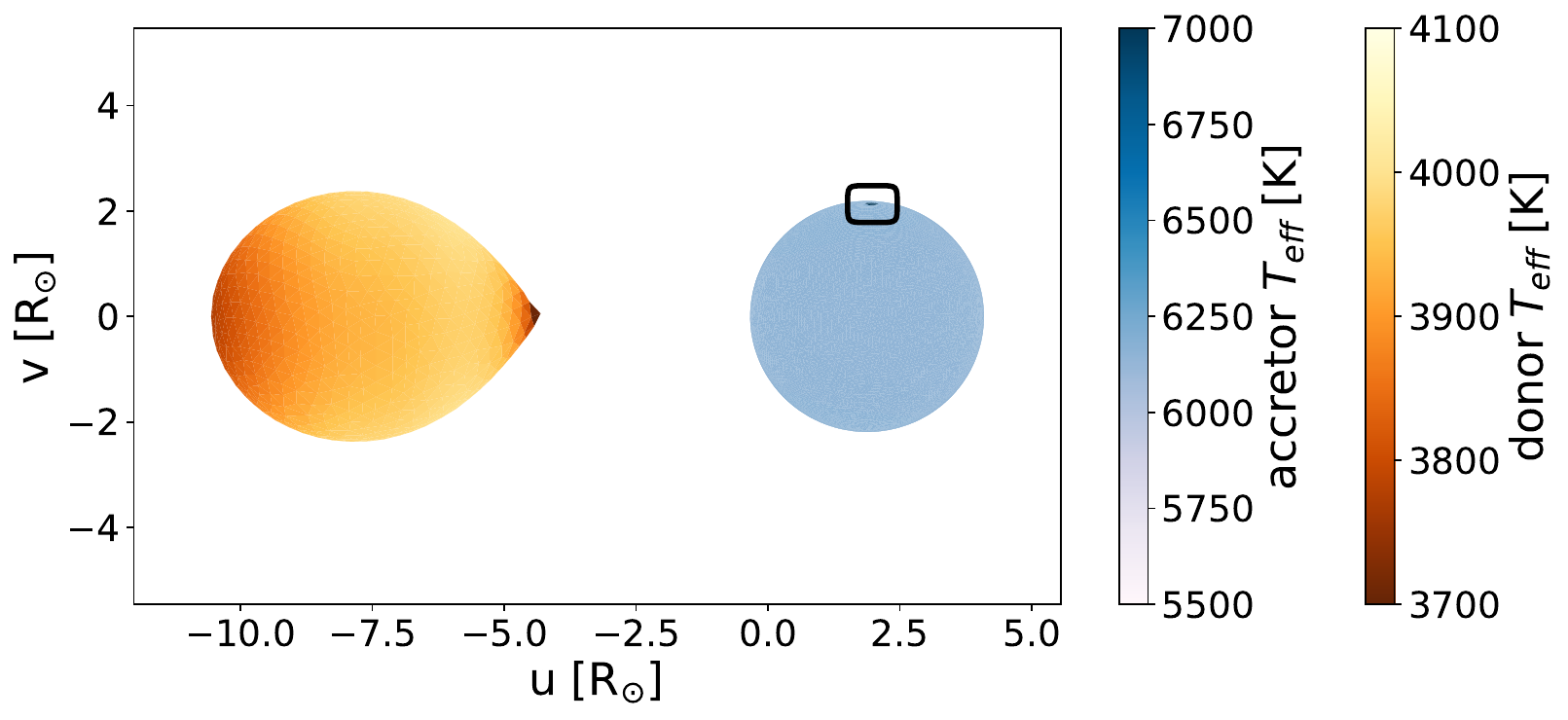}
\caption{Mesh showing the temperature distribution along the photospheres, including a spot. Top panel shows the unconstrained spotted model. Bottom panel shows the model with fixed \teff\ of the photosphere and of the spot. Note that in the latter case the spot is close to polar, highlighted by a black rectangle.
\label{fig:mesh}}
\end{figure*}

Given the availability of the light curves that show the system as eclipsing, as well as RVs for both stars, it is possible to perform a full fit of all of the stellar parameters. To do this, we use PHOEBE \citep{conroy2020}, which enables a comprehensive modeling of eclipsing binaries. In this section we describe different approaching to constructing such a model.

\subsection{Simple model}

In the process of modeling \eb, we consider 21 parameters in total. They include 9 stellar parameters: effective surface temperatures \teffa, \teffb, semi-mejor axis $a$, orbital period $P$, inclination $i$, mass ratio $q$, center-of-mass velocity $\gamma$, time of passage of the superior conjunction $t_0$, as well as the equivalent radius $R_1$ of the hotter star (accretor). The system is treated as semi-detached, with $R_2$ being set to the radius of the Roche lobe of the cooler star (donor). Due to the compactness of the orbit, circular orbit is assumed.

In addition to these stellar parameters we fit the third light contamination ($l_3$) for each of six light curves in different bandpasses that are available for the star, to accommodate flux contamination from the nearby field stars, as well as from the likely tertiary within the system. We also include passband luminosity for all six light curves, which is a nuisance variable recommended by PHOEBE.

We performed an initial manual exploration of the ranges of stellar parameters, as well as a preliminary optimization with the Nelder-Mead solver. Due to the combination of stellar parameters required by this system being outside of the set of the set of the atmospheres incorporated into PHOEBE, we treat atmospheres of both of the stars as blackbodies, with limb darkening computed using LDTK \citep{parviainen2015} from the larger set of PHOENIX atmospheres \citep{husser2013}.

We then fit the system using the emcee sampler. It was initialized with 250 walkers, and a chain length of 5,000 steps, which was sufficient for emcee to converge after 2,000 steps. In the distribution, we adopted $\gamma$ and $q$ and their uncertainties from \citet{kounkel2021}. We also imposed uniform priors on the system of 5500$<$\teffa$<$7500 K, 3500$<$\teffb$<$6500 K, 4$<\gamma<$8 \kms, 75$<i<90^\circ$, and 8.5$<a<$11.5 \rsun, to prevent emcee from walking away to implausible solutions early on in the chain, typically through ignoring sparse RVs, which delays the eventual convergence.

The resulting fitted parameters are also listed in Table \ref{tab:params}, and the evaluation of the model against the data is shown in Figure \ref{fig:withoutspot}. The fit well describes both the RV curve and the light curves, although it doesn't capture a slight asymmetry in the continuum flux in TESS. 

The third light contribution in TESS is comparable but somewhat larger than the a-priori estimate of 10\% from the neighboring field stars alone. Similarly, although Gaia has a significantly higher spatial resolution, it also requires a non-negligible $l_3$ of $\sim2$\%. This supports a presence of a significantly fainter third star in the system.

However, adopting these stellar parameter for the SED fit creates a significant discrepancy: the model significantly overestimates IR flux, and underestimates blue flux. This is partially due to the fact that temperatures of the two stars converge to a lower \teff\ than the ranges that are able to fit the SED. Forcing a tighter prior on \teff (such as at \teffa$>$6000 K) results in PHOEBE converging its \teff\ right at the edge of this prior.

The light curve modeling often has greater sensitivity to the ratio of the temperatures than the precise temperatures themselves, but even scaling \teff\ to the range in Figure \ref{fig:sedparams} cannot produce a good SED fit. The reason for this is because PHOEBE requires $R_2>R_1$ in all permutations, inconsistent with the SED derived ratio of the radii at the point of intersection of \teff\ ratios between the two approaches.

As such, the model requires an additional hotter source of light that would be able to reconcile the two approaches. This system is likely a tertiary, but the third star cannot be responsible for producing excess blue light: fitting the SED with three stars, holding two of them fixed at PHOEBE solutions results the third star being at least a factor of 3 times brighter than either of the stars in the eclipsing binary, which is inconsistent with $l_3$ estimates, and is unlikely to not have been reflected in the spectra of the system.

There is one additional discrepancy between the derived model and the data. Regardless of the adopted \teffa, the donor star is expected to have comparable flux to the accretor in NIR, but it is expected to be fainter at all wavelength. In contrast, APOGEE spectra suggest that the donor star to be more luminous in the H band. Even with the uncertainty driven by the third light, individual radii, as well as the ratio of \teffb/\teffa are strongly constrained by the available light curves, which in turn constrain the luminosity of stars at a given wavelength. The mostly likely explanation for this is that in performing cross-correlation, lower \teff\ template of the donor star was favored as at lower \teff\ various spectral lines are stronger. Because of \teff\ difference of $>$1000 K, the template mismatch resulted in a weaker cross-correlation strength for the hotter accretor star, despite the latter being marginally more bright in this wavelength range.

\subsection{Spotted model}

In order to reconcile parameters from SED and light curve fitting, we consider the effect of spots. To increase the fraction of the blue light in the system, the spot has to be hotter than the photosphere. As the system is semi-detatched, likely with active mass transfer, the shock from the accretion stream from the donor is able to excite a small fraction of the photosphere of the accretor \citep{kouzuma2019}. Because of this, the accretor is more likely to contain a necessary hot spot.

The hint of such a hot spot is present in the data. As previously mentioned, TESS light curve exhibits a slight asymmetry in its flux continuum that cannot be reproduced by a simple model. Furthermore, there is FUV excess in the SED fit that cannot be accounted by any combination of \teff\ and radii of the two stars: both of these factors can be explained by a presence of a spot.

Given that there is no additional periodicity other than the orbital period, it is likely that the rotation of the stars is tidally locked to their orbit. Both compactness of the orbit and the mass transfer can be responsible for this. With the orbital period of just 2.7 days and the radii of $\sim2.5$ \rsun, this results in rapid rotation, with velocities on the order of 40--50 \kms, consistent with the velocity spread seen in the CCF taking instrumental broadening into the account (Figure \ref{fig:ccf}).

To incorporate a spot in the SED fitting, we create a mesh in PHOEBE, which is then able to independently specify the temperature, surface gravity, and area within each of the facets of the mesh. SED is then constructed for all of the facets, and they are integrated together to produce the total flux for each star and the system as a whole.

We first explore the range of likely spot properties in the SED fitting. We randomly sample \teffa\ between 5500 and 6500 K, forcing \teffa=1.53\teffb, to match the previously derived temperature ratio, preserving PHOEBE derived radii from the spotless model. Spot sizes are drawn from 0 to 25 degrees, and temperature contrast from 1 to 2.5. Once again, \teffa$<$6000 K is strongly ruled out by this fit. Larger spots require spot temperature closer to \teff, smaller spots have to be significantly hotter (Figure \ref{fig:sedspot2}).

Similarly, keeping \teff\ fixed, we explore the dependence of the spot's size and contrast relative its position on the photosphere on the goodness of fit of both the light curves and the SED simultaneously by computing $\chi^2$ for each. The most likely solution is a compact equatorial spot $\sim$140$^\circ$ away from the point facing the donor (Figure \ref{fig:sedspot}. To better fit the UV excess, small spots are preferred, with temperature $>1.5$ hotter than the photosphere; this is consistent both for the SED only and the light curve only fit (Figure \ref{fig:sedspot1}).

We then initialize emcee with 250 walkers, adding temperature contrast, radius, and the position of the spot as parameters in addition to 21 parameters from the spotless model. The resulting parameters are listed in Table \ref{tab:params}, and the fit is shown in Figure \ref{fig:withspot}.

In large, the quality of the fit relative to the spotless model is very similar -- adding a spot has significantly improved the fit to the TESS light curve, however, it did not substantively alter any of the derived parameters compared to a spotless model, thus there isn't a substantive differences in the overall shape of the modelled light curves or the RV curves. As such, the quality of the resulting SED fit did not substantively improve. The model still overestimated IR flux because the derived \teffa\ was too small. And while the NUV flux is better fit, FUV flux is significantly overestimated, because the contrast ratio and/or the spot size appear to be too large to well fit the available photometry.

We subsequently attempted several permutations of the fitting procedure, placing different priors on \teffa\ and on the spot properties. Forcing hotter \teffa\ fixes discrepancy in IR, but not in UV, as the spot continued to be too luminous. Eventually we fixed these parameters to the values most likely required by the SED, namely \teffa=6150 K, spot size of 2.22$\circ$, and the temperature contrast of the spot of 1.8 (i.e., $T_{\rm spot}\sim$11,000 K). The parameters are listed in Table \ref{tab:params}, and the fit is shown in Figure \ref{fig:withspotfix}. The quality of the resulting fit to the light curve is slightly impacted in comparison to the version of the model in which \teff\ is unconstrained, but this allows a more self-consistent solution between both approaches. The distance to the system based on the SED fitting is 782$\pm$13 pc.

The favored spot location is different between the models with unconstrained and with fixed \teff. Regardless of \teffa\, model with unconstrained spot parameters favors a larger equatorial hot spot. Forcing the spot to be less luminous forces it to migrate north. The smaller it is, the more likely model would place it near the polar region. The final model with fully constrained spot temperature and size places it only 6.5$^\circ$ from the north pole. 

\subsection{Third Star} \label{sec:thirdstar}

\begin{figure}
\includegraphics[width={0.5\textwidth}]{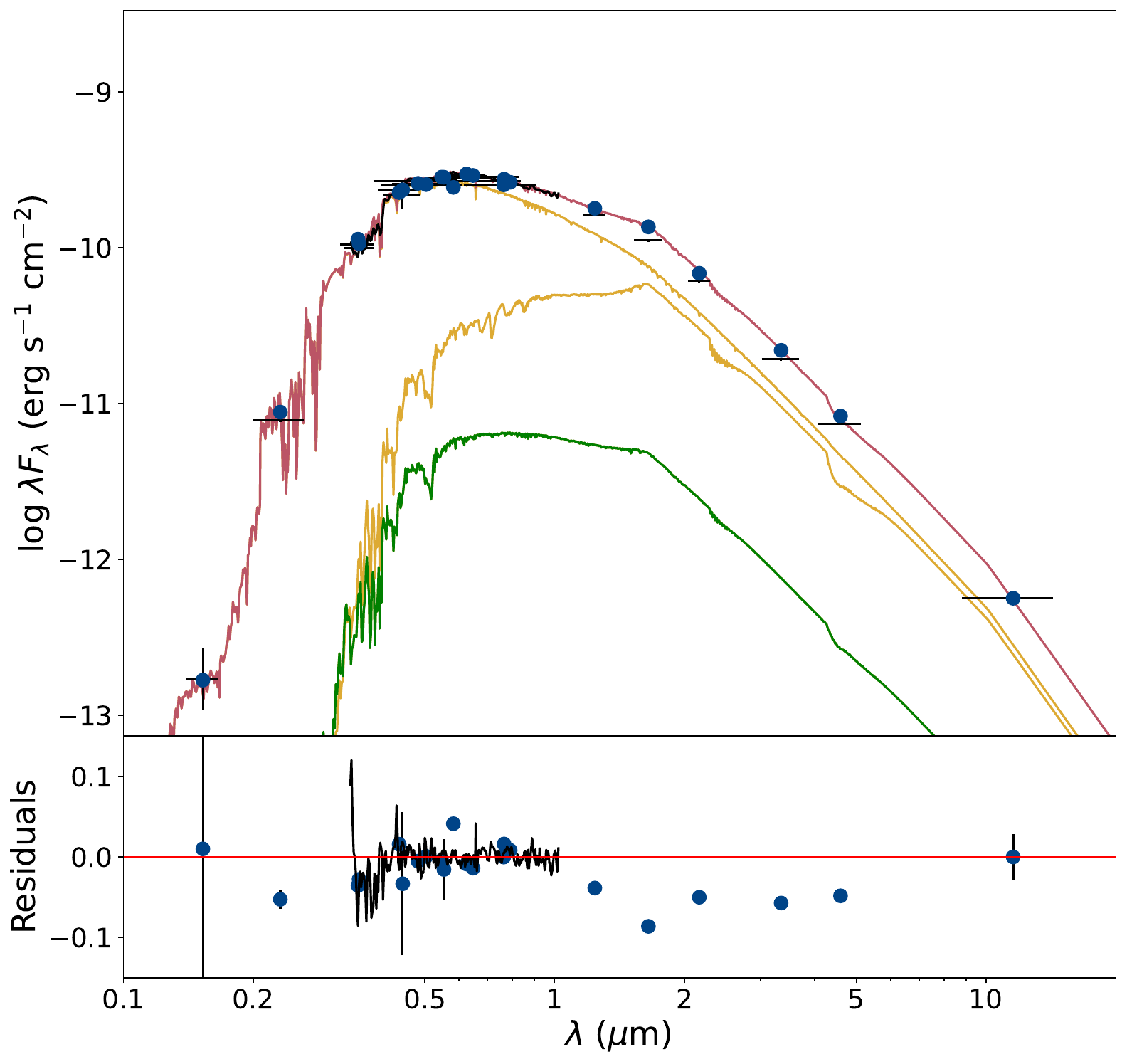}
\caption{SED fit to the system from Figure \ref{fig:withspotfix}, with the green curve shows the included teritary component consisting of a main sequence star with \teff=4600 K.
\label{fig:third}}
\end{figure}

As previously mentioned, very large RUWE, as well as non-negligible third light contribution in Gaia passbands is indicative of the presence of a tertiary star in the system. Since it amounts to only 2.6\% of the total flux at RP band, it does not significantly alter the SED fit, but it is likely to be a low mass main sequence star with \teff$\sim$4600 K with the corresponding radius of 0.66 \rsun, which matches the RP flux contribution (Figure \ref{fig:third}).

However, such a star significantly overestimates third light at G and BP passbands. It is possible that the model underestimates this contribution. Indeed, in all of the models, the fitted light curve somewhat underestimates the depth of the secondary eclipse in these bands, despite matching well in all of the other passbands.

This may be what has fundamentally led to the difficulty of deriving accurate \teff\ within PHOEBE, and requiring constraining it primarily from the SED fit. For example, constructing a PHOEBE model fit using only T, V, and g light curves, which are all contaminated by TIC 458723453 (and disregarding Gaia lighcurves) produces third lights of $l_{3,{\rm T}}$=8\%,  $l_{3,{\rm V}}$=0.1\%, and  $l_{3,{\rm g}}\sim$0 (much smaller than  $l_{3,{\rm T}}=$12\% and  $l_{3,{\rm V,g}}$=6\% in the model with the Gaia data). This model also had incorrect \teff, favoring \teffa$>$7000 K, which could not be reconciled with the SED fit.

It appears that PHOEBE struggles to accurately represent $l_3$ in cases where all of the passbands have some degree of contamination from a third body, favoring the lightcurve with the smallest contamination to be set at $l_3\sim$0, regardless of the absolute fraction, and this does impact the fitted \teff.

As such, $l_3$ in all passbands may still be underestimated by a few \%. In this case, the tertiary companion would likely be a main sequence 4600--5000 K star.

\section{Discussion} \label{sec:discussion}

\subsection{Accretion status} \label{sec:accr}

The current masses of the donor and the accretor star are 0.33 and 1.368 \msun. In the event of the conservative mass transfer and assuming that the masses two stars may have been initially similar, the minimum initial mas of the donor would be $>$0.85 \msun. As such, the donor has already lost well over half of its initial mass. We note that non-conservative mass transfer is common, including in the lower mass stars, with mass frequently lost due to winds, jet formation, outflows from the disk, etc \citep[e.g.,][]{woods2012,chen2020b,sun2021,sun2023}, thus, this is a lower limit. Given the proximity of the two stars, the onset of it would have been rapid following the expansion of the donor into a red giant. What is surprising, however, is that despite the history of substantial mass transfer in this system, it appears to still be active, as indicative through UV excess and the presence of a hot spot.

In young stellar objects that still actively accrete gas from protoplanetary disks, gas flows along the magnetic fields of a (proto)star, shocking the photosphere upon impact \citep[e.g.,][]{hartmann2016}. Such shock has been observed to create hot spot with the temperature contrast of $>$2 relative to photosphere \citep[$T_{\rm spot}\sim$ 9000 K in a young star with \teff$\sim$4000 K with $\dot{M}\sim1-2\times10^{-8}$ \msun yr$^{-1}$,][]{espaillat2021}. $T_{\rm spot}$ can reach upwards of $\sim$ 12,000 K \citep{de-sa2019}. This is not dissimilar to $T_{\rm spot}\sim$ 11,000 K that we observe in \eb.

In young stars, high density of the accretion flow tends to also produce spots with low filling factor of 0.1--1\% of the total size of the photosphere \citep{calvet1998a}, which can correspond to the spot sizes of a few degrees. On the other hand, lower density accretion columns will produce spots with larger filing factor \citep{ingleby2013}.

Hot spots have been previously observed in a number of close binaries \citep{kouzuma2019}. However, in systems with a similar configurations as \eb, i.e., a semi-detached binary in which a less massive star is overflowing its Roche lobe (SD2), contrast ratio of the spot to the photosphere is typically 1.1, and the spots themselves tend to be large, 10--30$^\circ$. As such, high $T_{\rm spot}$ relative to the temperature of the photosphere, as well as the compactness of the hot spot make \eb\ relatively unusual in its class. Most likely, this difference is caused by significantly higher $\dot{M}$ of \eb\ in contrast to a number of other semi-detached eclipsing binaries. This may be due to this system being caught very early on its evolution: after the bulk of mass has already been transferred between the two stars, but not yet at a point where it reaches a relative equilibrium in their respective masses.

\eb\ appears to be very similar to 2M17091769+3127589 \citep{miller2021a} in terms of the total mass of the system as well as the mass ratio. However, while donor of 2M17+31 is an evolved red giant with $R\sim$4 \rsun, \eb\ appears to have only recently begun climbing up the red giant branch. 

\subsection{Origin of discrepancy in the resulting PHOEBE \& SED models}

As previously discussed in Section \ref{sec:thirdstar}, the difficulty in determining accurate \teff\ in PHOEBE has most likely been caused by third light contamination, namely by underestimation of third light in G and BP bandpasses. However, even when holding \teff\ fixed in the model, there is a difference in the derived parameters for the spot between SED and PHOEBE fit. PHOEBE favors significantly more luminous spot, and, constraining the spot size and contrast results in a very different spot position.

In semi-detatched eclipsing binary systems, hot spots are most commonly equatorial \citep{kouzuma2019}, which is the location favored by the light curve in \eb. This gives greater credence to the PHOEBE derived spot properties, at least concerning its position. The luminosity of the spot may be skewed for the similar reason why \teff\ of the photosphere appears to be incorrect.

There is also a possibility of both SED-only and PHOEBE-only fit of the spot luminosity being accurate. The primary effect on photometry of a more luminous spot is in the FUV bandpass from GALEX. The only light curve with significant sensitivity to the spot is from TESS, as ASAS-SN has significantly lower signal-to-noise, and Gaia light curve is very sparse. GALEX and TESS observations have been separated in time by approximately a decade. During this time, there could have been an evolution in its accretion/spot properties, which may have increased the FUV luminosity, even though it is not captured in the available data. 

\subsection{Possible evolutionary history of this system}
\label{MESA models}

To explore the past and future of this system, we utilized the Modules for Experiments in Stellar Astrophysics (MESA; Version 12115; \citealt{Paxton2011, Paxton2013, Paxton2015, Paxton2018, Paxton2019, Jermyn2023}) binary module. The simulation adopts an initial metallicity of $Z = 0.019$. The model incorporates the same physics as described in Section 4 of \citet{miller2021a}. Our initial goal is to fit the MESA model to the fixed $T_{\rm eff}$ result from Table \ref{tab:params}, on the mass versus orbital period plane.

\begin{figure}
\includegraphics[width={0.5\textwidth}]{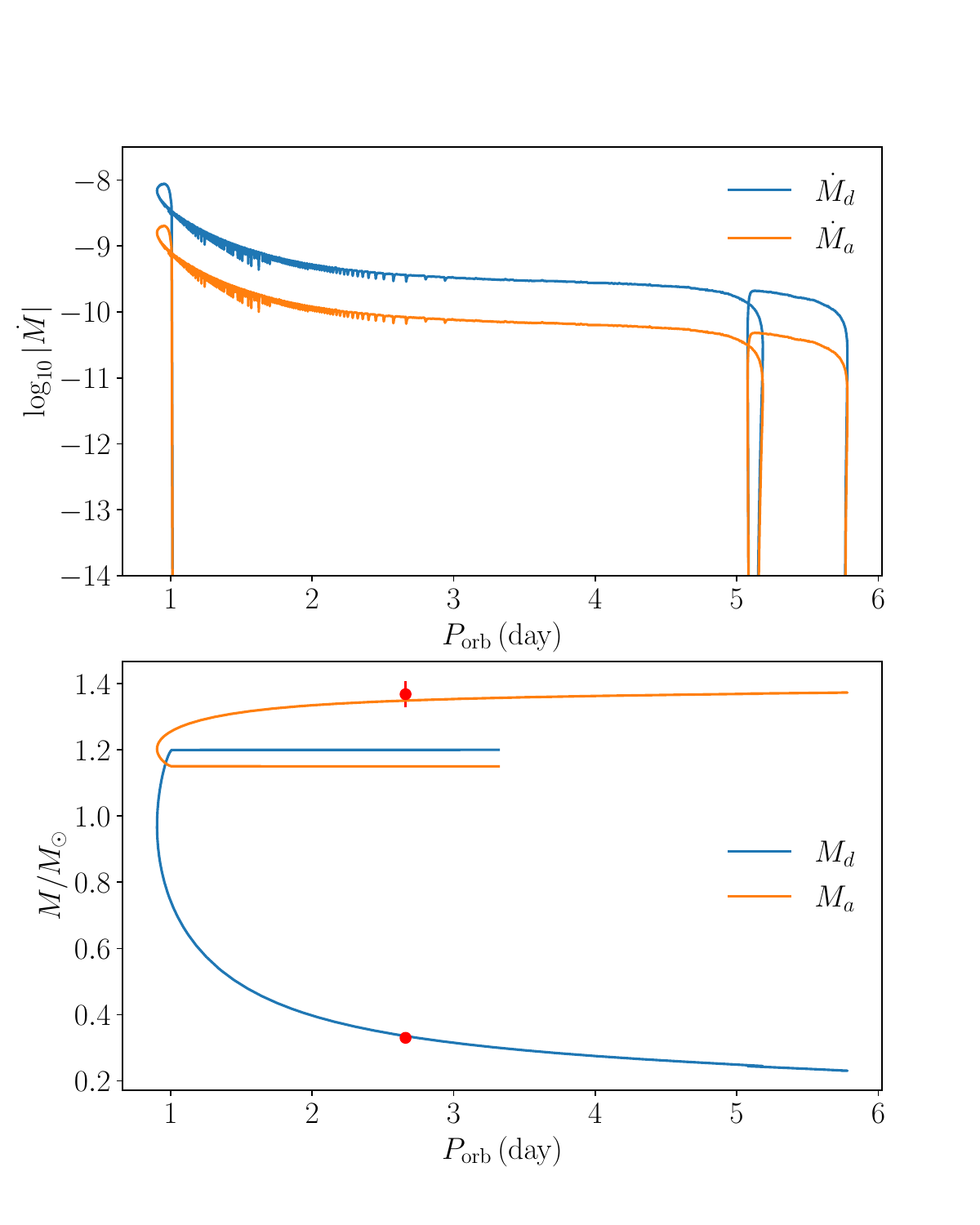}
\caption{The MESA model with initial donor mass $M_{\rm d} = 1.2\,M_{\odot}$, accretor mass $M_{\rm a}=1.15\,M_{\odot}$ and initial orbital period $P_{\rm orb}=3.32$ days, with a mass transfer efficiency of 23\%. In the upper subplot, the blue line represents the mass loss rate from the donor star, while the orange line represents the mass accretion rate of the accretor star. The lower panel illustrates the evolution of the donor (blue line) and accretor masses as a function of $P_{\rm orb}$ during the binary evolution. The error bar of the donor star mass measurement is smaller than the size of the data point, so it is not displayed on the plot.
\label{fig:Mdot_P}}
\end{figure}

Considering \eb\ has very similar features to 2M17091769+3127589, the initial parameters of the model were searched near the best-fit model of 2M1709. Figure \ref{fig:Mdot_P} shows the best fit model for \eb\ with an initial donor mass of $M_d = 1.2\,M_{\odot}$, an accretor mass of $M_a = 1.15\,M_{\odot}$, and an initial orbital period of $P_{\rm orb} = 3.32$ days (comparable with initial donor mass of $M_d = 1.2\,M_{\odot}$, $M_a = 1.11\,M_{\odot}$, and an initial orbital period of $P_{\rm orb} = 3.43$ days in 2M1709). To match the observed mass of the accretor star, the mass transfer efficiency is fixed at 23\% during the mass transfer (lower than 49\% in 2M1709). The system undergoes initial shrinkage during its evolution. The onset of mass transfer occurs at $P_{\rm orb} = 1$ day, when the donor star has already exhausted its central hydrogen. Around $P_{\rm orb} = 0.9$ day, with $M_{\rm d} = 0.97\,M_{\odot}$ and $M_{\rm a} = 1.2\,M_{\odot}$, the system begins to expand in $P_{\rm orb}$ and eventually reaches its current binary configuration, with a mass loss rate from the donor star of $\dot{M} \sim 10^{-9.5}\,M_{\odot}/{\rm yr}$. At the same time, the accretor star is in the subgiant phase, as indicated by both the modeled and observed radius and \teff. At the end of the simulation, the donor star evolves into a 0.23 $M_{\odot}$ helium pre-white dwarf, while the accretor star becomes a 1.36 $M_{\odot}$ red giant star. The simulation terminated when the accretor fills the Roche lobe radius, resulting in a contact binary. Most likely, after that point the system will evolve into a planetary nebula, as there is not enough mass to produce a Type 1a supernova.

\begin{figure}
\includegraphics[width={0.5\textwidth}]{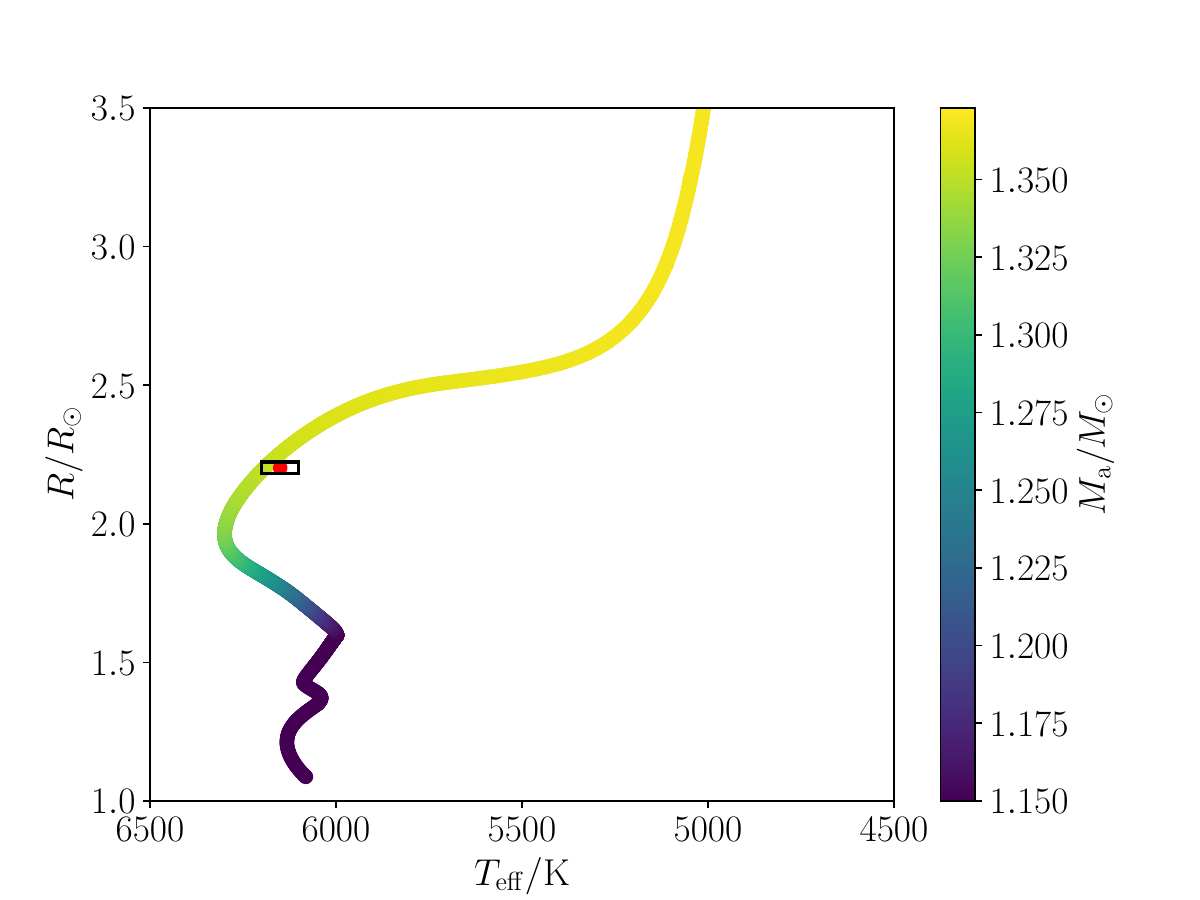}
\includegraphics[width={0.5\textwidth}]{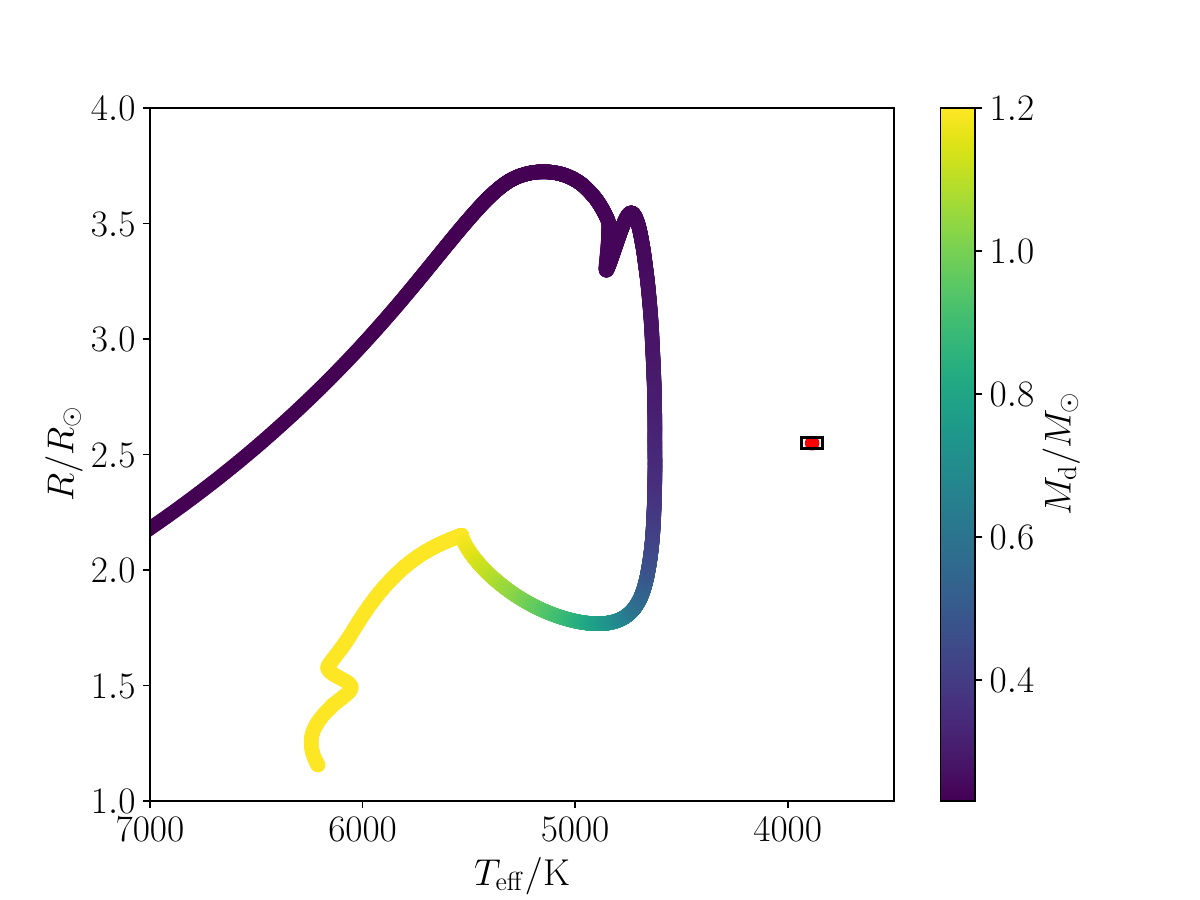}
\caption{Top: The evolutionary track of the fitted MESA model for the accretor star on the radius versus \teff\ plane. The data represented as red dot is from the third column of Table \ref{tab:params}. The size of the box correspond to the size of the error bars, while the color represents the mass of the accretor throughout its evolution. Bottom: same as above but for the donnor star.
\label{fig:R_Teff_accr}}
\end{figure}

This model explains the current masses, orbital period of the system, and the \teff\ and radius of the accretor star. Figure \ref{fig:R_Teff_accr}a illustrates the evolutionary track of the accretor star in terms of radius and $T_{\rm eff}$. The model begins at the zero-age main-sequence stage of the stars, characterized by $(T_{\rm eff},\, R) = (6070{\rm K},\,1.1 R_{\odot})$. Mass accretion commences during the accretor's post-main-sequence evolution at $(T_{\rm eff},\, R) = (6000{\rm K},\,1.6 R_{\odot})$. At the observed $P_{\rm orb} = 2.66$ days, the evolutionary track passes through the box indicating the error bar of the measurements, corresponding to a mass of 1.35 $M_{\odot}$.

\subsection{\teff\ of the donor star}

The temperature of the donor star is significantly hotter than observed, with a difference of approximately 800 K, as depicted in Figure \ref{fig:R_Teff_accr}b. The evolutionary track of the donor star commences at $(T_{\rm eff},\, R) = (6200{\rm K},\,1.12 R_{\odot})$. The onset of mass transfer takes place at $(T_{\rm eff},\, R) = (5550{\rm K},\,2.12 R_{\odot})$. The model indicates a case B mass transfer scenario. The discrepancy of 800 K between the model and data may not be solely attributed to model fine-tuning issues. We explored alternative parameter spaces, such as using a lower mass donor, but the \teff\ during mass transfer remains hotter than observed. None of the MESA models we have explored were able to more optimally fit the parameters of the system.

It is unclear why there is such a significant discrepancy in the donor's $T_{\rm eff}$. One of the possibilities might have been due to the fast rotation of the donor produce gravitational darkening that caused the star to appear cooler in comparison to a non-rotating model. This, however, is unlikely, as the difference of $\sim$1000 K would require rotation rate of 0.9 times the critical (\citealt{Paxton2019}, see Figure 38); the rotation velocity of $\sim$60 \kms\ observed for this system is significantly smaller.

The donor star may be a sub-sub-giant (SSG), which are often identified though being luminous but cooler than the red giant branch stars \citep{mathieu2003}. Much about their formation is still uncertain, and evolutionary models still struggle to fully reproduce their evolutionary history. SSGs may often have large cool spots on their surface which may partially explain \teff\ discrepancy \citep{stassun2023} -- the light curves for \eb\ show no evidence of these cool spots, as a single hot spot is sufficient for accounting for the out-of-eclipse variability. Nonetheless, the data may not necessarily rule out a possibility of them being present. Given the similarity of \eb\ to 2M1709, it is possible that the donor star being significantly cooler than what is predicted by the model is a transitory stage, and as it expands further, it may return to its expected position on the red giant branch.

Further resolving the temperature discrepancy of the donor star would be the focus of the future work. We note, however, given the strong presence of the accretion signatures that were discussed in Section \ref{sec:accr}, combined with an imprecise fit, it is possible that $\dot{M}$ in the model is underestimated due to the limitations of the evolutionary model being only 1D. 

\section{Conclusions} \label{sec:conclusions}

Through analyzing light curves from TESS, Gaia, and ASAS-SN, as well as radial velocities from APOGEE, we derive stellar and orbital properties of a semi-detatched eclipsing binary system, \eb. This system has a significant history of mass transfer, with the donor star transferring well over half of its mass onto the accretor, resulting in the present day masses of 0.33 \msun\ and 1.37 \msun respectively. Despite his, the system still exhibits active accretion, as is evident by a very compact and very luminous hot spot on the surface of the accretor.

There have been some challenges in deriving accurate \teff\ for the individual stars due to third light contamination, both from a low mass main sequence companion orbiting around the binary, as well as a nearby field star TIC 458723453. Deriving accurate \teff\ for both stars required an iterative process of both the light curve and the SED modelling, as the latter can place firmer constraints on the allowable range of \teff. Incorporating SED fitting into the light curve fitting routines such as PHOEBE could yield more self-consistent models.

We note that the donor star in \eb\ appears to be a sub-sub-giant. However, this system appears to be a twin to 2M17091769+3127589 \citep{miller2021a}, potentially starting out with near-identical initial conditions, but significantly less evolved. Further analysis of these two systems may shed more light on the formation and evolution of SSGs.

\section*{Acknowledgements}

M.S. thanks Seth Gossage for the valuable discussions on stellar rotation.

M.S. acknowledges support through grant GBMF8477.

\section*{Data Availability}

RV data has been obtained from \citet{kounkel2021}; APOGEE spectra for this source are publicly available \citep{abdurrouf2022}. Gaia light curves are included in Gaia DR3 \citep{gaia-collaboration2022a}. TESS data are hosted on \url{https://archive.stsci.edu/}, and ASAS-SN data on \url{asas-sn.osu.edu/}. All of the resulting data products and models can also be made available upon request.



\bibliographystyle{mnras}
\bibliography{main.bbl}





\bsp	
\label{lastpage}
\end{document}